\documentclass[aps,nofootinbib,notitlepage,superscriptaddress,twocolumn,10pt,prd]{revtex4-1}

\usepackage{graphicx}
\usepackage{amsmath,amssymb}
\usepackage{hyperref}
\usepackage{braket}
\usepackage{subfigure}
\usepackage{float}
\usepackage[dvipsnames]{xcolor}
\usepackage{soul}
\usepackage{rotating}
\usepackage{multirow}
\usepackage{mathtools}
\usepackage[normalem]{ulem}
\usepackage{comment}
\usepackage{cprotect}
\usepackage{xspace}

\usepackage{makecell}

\usepackage[margin=0.75in]{geometry}

\hypersetup{
	colorlinks=true,
	linkcolor=blue,
	citecolor=blue,
	urlcolor = black,
}

\usepackage{xcolor}

\def\apj{Astrophys.\ J.}

\def\mnras{Mon.\ Not.\ Roy.\ Astron.\ Soc.}

\def\prd{Phys.\ Rev.\ D}

\def\sptpol{SPTpol}
\def\planck{{\it Planck}}

\newcommand{\rdrag}{\ensuremath{r_{\rm d}}}
\newcommand{\QUDM}{\ensuremath{Q_\textrm{\tiny UDM}}}
\newcommand{\BAO}{{\textrm{\tiny BAO}}}
\newcommand{\zBAO}{z_{\textrm{\tiny BAO}}}
\newcommand{\thetapar}{\theta_\parallel}
\newcommand{\thetaperp}{\theta_\perp}
\newcommand{\BB}{BAO+BBN\xspace}
\newcommand{\BBS}{BAO+BBN+\allowbreak\sptpol\ lensing\xspace}
\newcommand{\BBP}{BAO+BBN+\allowbreak\planck\ lensing\xspace}
\def\be{\begin{equation}}
\def\ee{\end{equation}}
\def\ba{\begin{eqnarray}}
\def\ea{\end{eqnarray}}
\def\({\left(}
\def\){\right)}

\newcommand{\refsec}[1]{Sec.~\ref{sec:#1}}
\newcommand{\reffig}[1]{Fig.~\ref{fig:#1}}
\newcommand{\refeqn}[1]{Eq.~(\ref{eqn:#1})}

\begin{document}
	
\title{Hubble constant difference between CMB lensing and BAO measurements}

\author{W.~L.~Kimmy Wu}
\affiliation{Kavli Institute for Cosmological Physics, University of Chicago, Chicago, Illinois 60637, U.S.A}

\author{Pavel Motloch}
\affiliation{Canadian Institute for Theoretical Astrophysics, University of Toronto, M5S 3H8, ON, Canada}

\author{Wayne Hu}
\affiliation{Kavli Institute for Cosmological Physics, Department of Astronomy \& Astrophysics, Enrico Fermi Institute, University of Chicago, Chicago, Illinois 60637, U.S.A}

\author{Marco Raveri}
\affiliation{Center for Particle Cosmology, Department of Physics and Astronomy, University of Pennsylvania, Philadelphia, PA 19104, USA}

\begin{abstract}	
We apply a tension metric $\QUDM$, the update difference in mean parameters, to understand the source of the 
 difference in the measured Hubble constant $H_0$ inferred with cosmic microwave background lensing measurements from the \planck\ satellite ($H_0=67.9^{+1.1}_{-1.3}\, \mathrm{km/s/Mpc}$) and from the South Pole Telescope  ($H_0=72.0^{+2.1}_{-2.5}\, \mathrm{km/s/Mpc}$)
when both are combined with baryon acoustic oscillation (BAO) measurements with priors on the baryon density (BBN).
$\QUDM$ isolates  the relevant parameter directions for tension or concordance where the two data sets 
are both informative, and aids in  the identification of subsets of data that source the observed tension. 
With $\QUDM$, we uncover that the difference in $H_0$ originates from differences between \planck\ lensing and \BB data, 
at probability-to-exceed of 6.6\%.
Most of this mild disagreement comes from the galaxy BAO measurements parallel to the line of sight
in the redshift range $0.2 <  z < 0.75$.
The redshift dependence of the parallel BAOs pulls  both the matter density $\Omega_m$ and $H_0$ high  in $\Lambda$CDM,
but these parameter anomalies are usually hidden when the BAO measurements are combined with other cosmological data sets with much stronger $\Omega_m$ constraints.
\end{abstract}
	
\maketitle

\section{Introduction}
The standard cosmological model $\Lambda$CDM is extremely successful in describing observations over
a wide range of scales and redshifts: from the cosmic microwave background (CMB)
to the expansion of the universe today. 
However, increasingly precise measurements of cosmological parameters obtained in the past
several years uncovered mild to strong tensions between different data sets. 
Most notably, \planck\ infers the Hubble constant to be $H_0=67.36 \pm 0.54 \, \mathrm{km/s/Mpc}$ under $\Lambda$CDM \cite{Aghanim:2018eyx}
while the Cepheid-calibrated Type Ia Supernovae from SH0ES gives 
$H_0 =74.03 \pm 1.42 \, \mathrm{km/s/Mpc} $~\cite{riess2019} (see \cite{Freedman:2019jwv,Freedman:2020dne,Wong:2019kwg,Yuan:2019npk,Pesce:2020xfe} for other  measurements).
Such tensions between different data sets could
suggest the need to extend the $\Lambda$CDM model
to accommodate the observations or, alternatively, the existence of unmodeled systematics
in the data sets~\cite[e.g.][]{Bernal:2016gxb, Karwal:2016vyq, Mortsell:2018mfj, Aylor:2018drw, Poulin:2018cxd, Agrawal:2019lmo, Lin:2019qug, Kreisch:2019yzn, Raveri:2019mxg, Knox:2019rjx}. 
It is thus very important that we find independent measurements that can clarify the source(s) 
of the current $H_0$ tension.

One such example has been reported recently for the data sets that result from measurements of the 
weighted gravitational potential integrated along the line of sight (CMB lensing)
from the \planck\ satellite and the South Pole Telescope (\sptpol),
baryon acoustic oscillation (BAO) parallel and perpendicular to the line of sight in galaxy surveys,
and the baryon density inferred from the deuterium abundance ($D/H$) measurements (denoted by BBN). 
In~\cite{Bianchini:2019vxp}, the values of $H_0$ inferred from the \BBP and the \BBS
data sets are
$67.9^{+1.1}_{-1.3}\, \mathrm{km/s/Mpc}$ and
$72.0^{+2.1}_{-2.5}\, \mathrm{km/s/Mpc}$ 
respectively.  
This mild discrepancy is intriguing since it is reminiscent of the tension between 
the SH0ES  vs.~\planck\  CMB power spectra measurements above.

In addition, the constraints on the lensing amplitude as captured by the parameter
combination $\sigma_8\Omega_m^{0.25}$ between the two lensing data sets are completely
consistent.   This presents somewhat of a puzzle since the mild disagreement in $H_0$ appears 
through adding \BB to the otherwise consistent 
lensing data sets.
Therefore, we set out to investigate the underlying driver(s) of the differences in the inferred $H_0$.

To do this, we apply a tension metric developed in~\cite{raveri18}
to quantify tension between \BBP and \BBS. 
We find that the difference in $H_0$ between the two 
 is driven by the different inferences from the shape of the \planck\ lensing spectrum and \BB.
Specifically, it is the line-of-sight BAO measurements 
that pull $H_0$ high.
This preference is ordinarily hidden when the line-of-sight BAO measurements are combined with other cosmological data sets, since  it 
requires a high matter density in $\Lambda$CDM which is strongly ruled out by these other data sets. 

This paper is organized as follows: In~\refsec{tensions} we summarize the
tension metric we use, before presenting the data sets used in this work in~\refsec{data}.
In~\refsec{lensbao}, we isolate and quantify the tension between
the \BBS and the \BBP data sets {and} show that it originates from 
the \planck\ lensing and the parallel BAO measurements.
In~\refsec{otherdata}, we show that values of cosmological parameters preferred by
the parallel BAO measurements are strongly ruled out by other cosmological data sets.
We discuss the results and conclude in~\refsec{discuss}.

\section{Quantifying tensions}
\label{sec:tensions}

To quantify tensions between uncorrelated data sets, we use the update difference-in-mean (UDM) statistic defined in~\cite{raveri18}.
This statistic compares the mean parameter values from a data set $A$ alone,
$\theta^A_{\alpha}$, with their ``updated'' values after adding another data set $B$ to
$A$, $\theta^{A+B}_\alpha$. The index $\alpha$ here enumerates the individual
 parameters.
Specifically, this statistic computes the square of the difference between the mean
parameter values of the two sets
\be
\Delta \bar\theta_\alpha = \bar\theta^A_\alpha - \bar\theta^{A+B}_\alpha ,
\ee 
in units of its  covariance $C_\Delta$, 
\be
\label{eq:QUDMtot}
	\QUDM^{\rm tot} = \sum_{\alpha,\beta} 
		\Delta \bar\theta_\alpha
		\(C^{-1}_\Delta\)_{\alpha\beta}
		\Delta \bar\theta_\beta .
\ee
$C_\Delta$ is inferred from  the covariances of $A$ and $A+B$ as
\be
\label{eq:CDelta}
	C_\Delta = C^A - C^{A+B}.
\ee
To quantify the tension, we use the fact that if  the $A$ and ${A+}B$ parameter posteriors
 are Gaussian distributed
and drawn from a  self-consistent model,  $\QUDM^{\rm tot}$ is chi-squared distributed with the number of parameters measured by both $A$ and $B$
as the degrees of freedom.

There are two advantages  of using $\QUDM^{\rm tot}$ as opposed to the simpler difference-in-mean statistic.
The first is practical.  If the parameter posterior for $B$ is highly non-Gaussian due to weak constraints or degeneracies, then
the squared difference-in-mean is also far from chi-squared distributed, making its significance
difficult to quantify.   
However, if the parameter posterior for  $A$ is more Gaussian, then the posterior of the
combined data set $A+B$ is also more Gaussian and $\QUDM^{\rm tot}$ is closer to chi-squared distributed.
The second advantage is the ability to pre-identify parameter
directions in
which the combination of  $A$ and $B$ improves the errors over $A$ or $B$ 
individually and hence can exhibit interesting tension or confirmation bias.
This is in contrast to defining the parameter space of investigation by inspecting the 
parameter differences in mean {\it a posteriori}, e.g.\ by picking the most discrepant directions.

An effective method to isolate these directions is to identify 
the Karhunen-Lo\`{e}ve (KL) eigenmodes of the covariance matrices  \cite{raveri18}.   These eigenmodes 
$\phi^a_\alpha$
are
the solutions to the  generalized eigenvalue problem
\be
\sum_\beta C^A_{\alpha \beta} \phi^a_\beta
=
\lambda^a \sum_\beta C^{A+B}_{\alpha \beta} \phi^a_\beta ,
\ee
and are normalized so that 
\be
	\sum_{\alpha\beta} \phi_\alpha^a C^{A+B}_{\alpha\beta}\phi^b_\beta
	=
	\delta^{ab} .
\label{eqn:KLnorm}
\ee
The parameters in the KL basis
\be
	p^a = \sum_\alpha \phi_\alpha^a  \theta_\alpha
\ee
are uncorrelated for both $A$ and $A+B$ with 
variance $\lambda^a$ and $1$, respectively.
From Eqs.~(\ref{eq:QUDMtot}) and (\ref{eq:CDelta}),
\be
	\QUDM = \sum_a \frac{\(\Delta p^a\)^2}{\lambda^a - 1} .
	\label{eq:KLQUDM}
\ee
If the sum is over all KL modes then $\QUDM=\QUDM^{\rm tot}$.  However, to isolate the
directions of interest, 
we restrict this sum to eigenvalues 
\begin{equation}
0.1 <  \lambda^a-1  < 100.
\label{eq:KLrange}
\end{equation}
Notice that this selection does not involve the actual values of the difference in means, only the expected
ability of $B$ to update $A$ and vice versa.  
For cases where $\lambda_a \approx 1$, data set $B$ does not update the constraints of $A$ appreciably whereas
for $\lambda_a \gg 1$, data set $A$ itself becomes irrelevant and cannot update $B$.  In the former case, there are
also numerical problems due to the MCMC sampling of the posteriors.  
This selection also
covers cases where there are nuisance parameters that are constrained by only one of $A$ or $B$.

 With only the interesting directions in the parameter space
retained, we can then determine the  significance of their associated difference in means
by noting that $\QUDM$ is  chi-squared distributed with the number of remaining parameter directions
as the degrees of freedom.   From this point forward we refer to $\QUDM$ as defined by Eqs.~(\ref{eq:KLQUDM}) and (\ref{eq:KLrange}) as
 the update difference-in-mean statistic.

We will also be interested in how much information each KL eigenmode
contributes to constraining individual cosmological parameters $\theta_\alpha$.
Recall that 
 the Fisher information matrix is the inverse of the parameter covariance matrix $F_{\alpha\beta} =(C_{\alpha\beta})^{-1}$ and each diagonal entry corresponds to the inverse variance of the parameter if all other parameters 
 are held fixed.
Using \refeqn{KLnorm}, we can express 
this Fisher information of data set $A$  for the parameter $\theta_\alpha$  as
\be
	F_{\alpha\alpha} = \sum_a
	F^a_{\alpha\alpha}\ =   \sum_a \phi^a_{\alpha}  \phi^a_{\alpha} / \lambda^a .
\ee
The Fisher information of data set $A+B$ is the same 
expression with $\lambda^a\rightarrow 1$, but this will not be needed in our analysis below.

The fractional Fisher information $F^a_{\alpha\alpha}/F_{\alpha\alpha} \in [0, 1]$
parameterizes how important  KL mode $a$  is in constraining the cosmological parameter
$\theta_\alpha$, where low values mean that dropping this mode does not significantly affect 
its constraints.

When considering correlated data sets,  in particular  the internal consistency of
parallel and perpendicular BAO measurements, we use  the generalization of the above
discussion as in~\cite{Raveri:2019gdp}. Specifically, we duplicate the parameter space of
the model, and fit the joint data set with one copy of parameters controlling the theory
prediction for the first part of the joint data set (e.g.~parallel BAOs) and the other copy controlling the
theory prediction for the second part (e.g.~perpendicular BAOs).   We then assess the confidence intervals of the
difference
in these two parameter sets by sampling its posterior.  
Because we fit the joint data set, the correlations are properly accounted for.
This technique also has the benefit of applying to non-Gaussian posterior 
distributions.

\section{Data}
\label{sec:data}

The data sets we investigate in~\refsec{lensbao} include
\planck\ 2018 lensing~\cite{planck2018lens}, \sptpol\ lensing~\cite{wu2019},
BAO from SDSS DR12 BAO consensus sample~\cite[DR12,][]{Alam:2016hwk}, Main Galaxy Sample~\cite[MGS,][]{sdssmgs}, 
and the 6dF Galaxy Survey ~~\cite[6dF,][]{sdss6df}, and baryon density prior motivated by~\cite{cooke17} 
using $D/H$ measurements.
In~\refsec{otherdata} we compare BAO constraints with parameter constraints
from the \planck\ temperature and polarization power spectra \cite{Aghanim:2019ame}
and from the Pantheon supernova sample \cite{Scolnic:2017caz}.

The applicability of the 
$\QUDM$ statistic hinges upon the data sets being uncorrelated.
For the data sets we are considering, 
the BAO measurements are uncorrelated with the lensing measurements.
While the \planck\ and the \sptpol\ measurements have partial sky overlap, the overlap is very small ($\sim 1\%$)
and the angular scales of overlap are small as well (with \planck's $L=[8,400]$ and \sptpol's $L=[100, 2000]$).
Therefore, the lensing data sets are nearly uncorrelated.
Consequently $\QUDM$ is appropriate for quantifying tension between our data sets.  We only require
the parameter duplication generalization discussed in the previous section to quantify tension between
correlated subsets of a given data set (e.g.~BAO).

In all cases, we use {\sc CosmoMC}~\cite{Lewis2002} to sample the posteriors of these data sets.
We impose the following priors for $\Lambda$CDM parameters when sampling:
uniform priors for the cold dark matter density $\Omega_ch^2 = [0.001, 2.99]$,
initial curvature power spectrum  amplitude ${\rm ln}(10^{10} A_s) = [1.61, 3.91]$,
and the effective angular sound horizon scale $\theta_{\rm MC} = [0.5, 10]$.
We assume 
Gaussian priors (mean, $\sigma$) for the initial spectrum tilt $n_s: (0.96, 0.02)$ and the baryon density
$\Omega_bh^2: (0.0222, 0.0005)$ 
 (the latter representing the $D/H$ measurements/BBN data) and we fix the
optical depth to recombination $\tau$ to 0.055.
To draw the contour plots, we use GetDist \cite{Lewis:2019xzd}.

\section{\sptpol\  and Planck lensing vs. \BB}
\label{sec:lensbao}
In this section, we start with identifying the key parameter directions
 that contribute to the apparent disagreement
between \sptpol\ lensing and \planck\ lensing
when both data sets are combined with \BB using 
the update difference in mean statistic of~\refsec{tensions}. 
Upon finding that the apparent disagreement is not between 
\sptpol\ and \planck\ lensing, but rather between \planck\ lensing and the
parallel BAO measurements, we then focus on and quantify tension between those
measurements.

\subsection{Parameters}

The BAO+BBN+lensing data sets depend on 5 out of the 6 $\Lambda$CDM parameters (they are not
sensitive to $\tau$).
In order to obtain more Gaussian covariance matrices, we perform our investigations in the
parameter space that is native to the BAO and CMB lensing measurements.
This way, it is also easier to interpret the influence of each. 
Specifically, we work in the parameter basis 
\begin{equation}
\theta = \left[ \thetaperp,\thetapar,  \sigma_8\Omega_m^{1/4}, \Omega_bh^2, n_s \right].
\end{equation}
Here 
\begin{align}
\thetaperp & =  D_M(z_\BAO)\frac{ r_{\rm fid}}{\rdrag} ,\nonumber\\
\thetapar &= H(z_\BAO) \frac{\rdrag}{r_{\rm fid}},
\end{align}
where
$D_M(z)$ is the comoving angular diameter distance to redshift $z$, $H(z)$ is the
expansion rate at this redshift, $\rdrag$ is the comoving BAO scale (the sound horizon at the end of the Compton drag epoch), 
$r_{\rm fid} \equiv 147.78 \, {\rm Mpc}$ is the fiducial $\rdrag$,
 and $\sigma_8$ is the
root mean square of the linear matter density fluctuations at the $8 h^{-1}$ Mpc scale.
We choose  $z_\BAO= 0.61$, one of the DR12 points,
 but other
choices of $z$ within the DR12 range do not qualitatively affect our results.

The first three parameters are the most relevant to this work and may be interpreted as
a perpendicular BAO, parallel BAO, and CMB lensing amplitude parameter. 
Given that under $\Lambda$CDM $H_0 D_M(z)$ and $H(z)/H_0$ are functions of $\Omega_m$ alone, 
the first two parameters  span the same space as do $\Omega_m$ and
 $H_0 \rdrag$.
$\Omega_b h^2$ is constrained mainly by the BBN data and $n_s$
is a nuisance parameter that is constrained by the prior given in~\refsec{data}.

The BAO measurements do not depend on the lensing amplitude parameter $\sigma_8
\Omega_m^{1/4}$, while the shape of the lensing power spectrum within $\Lambda$CDM does depend
on the BAO parameters, in this case mainly supplying extra information on $\thetapar$.
The difference between SPTpol and \planck\ lensing  
can be attributed to the fact that the 
former mainly constrains the amplitude of the  lensing power spectrum whereas the latter
constrains both the amplitude and shape \cite{Bianchini:2019vxp}.

\begin{figure}
\begin{center}
\includegraphics[width=0.48\textwidth]{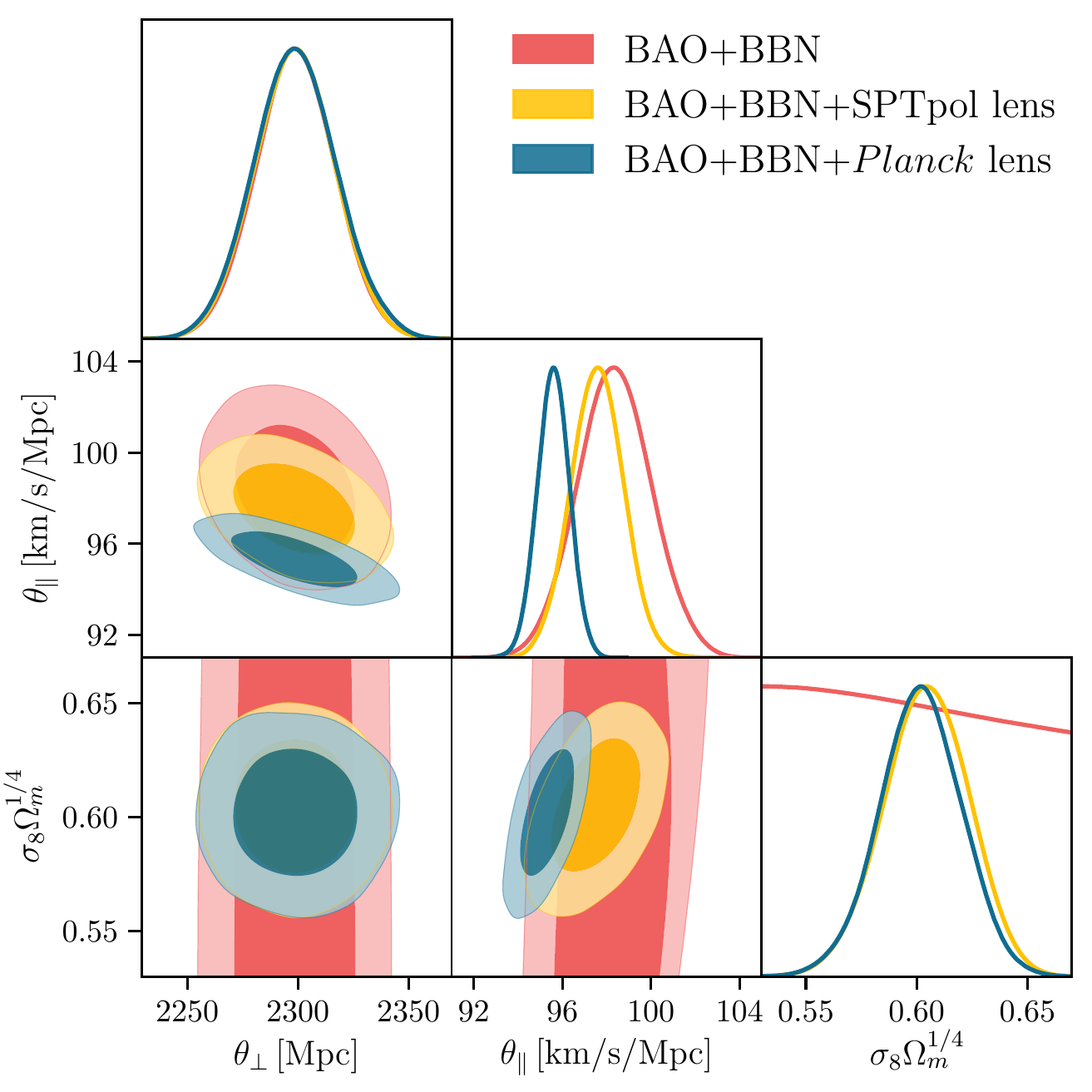}
\caption{The posterior distributions of the data sets
\BB, \BBS, \BBP in 
the parameter basis $[ \thetaperp,\thetapar,  \sigma_8\Omega_m^{1/4} ]$
which is native to the BAO and CMB lensing measurements.
The posteriors for the combined BAO+BBN+\planck\ lensing+\sptpol\
lensing data set are not shown for clarity as they are qualitatively close to
those from \BBP.
}
\label{fig:lens_bao_dm_h}
\end{center}
\end{figure}

We  can see some of these properties in the posterior distributions of these three parameters
shown in~\reffig{lens_bao_dm_h}.   In this space, the difference between \BBS
and \BBP  is  confined to the 
parallel BAO parameter $\thetapar$.  
From the \BB result, we can see  the shift in this parameter is already present
without the \sptpol\ lensing data, albeit with larger uncertainties.
Another notable observation is that the lensing amplitude parameter constrained
by \sptpol\ lensing and \planck\ lensing appears to agree too well with each other. 
Finally, the perpendicular BAO parameter $ \thetaperp$ is mainly constrained by the BAO data themselves
and the addition of either lensing data set does not change its posterior appreciably.

\begin{figure}
\begin{center}
\includegraphics[width=0.48\textwidth]{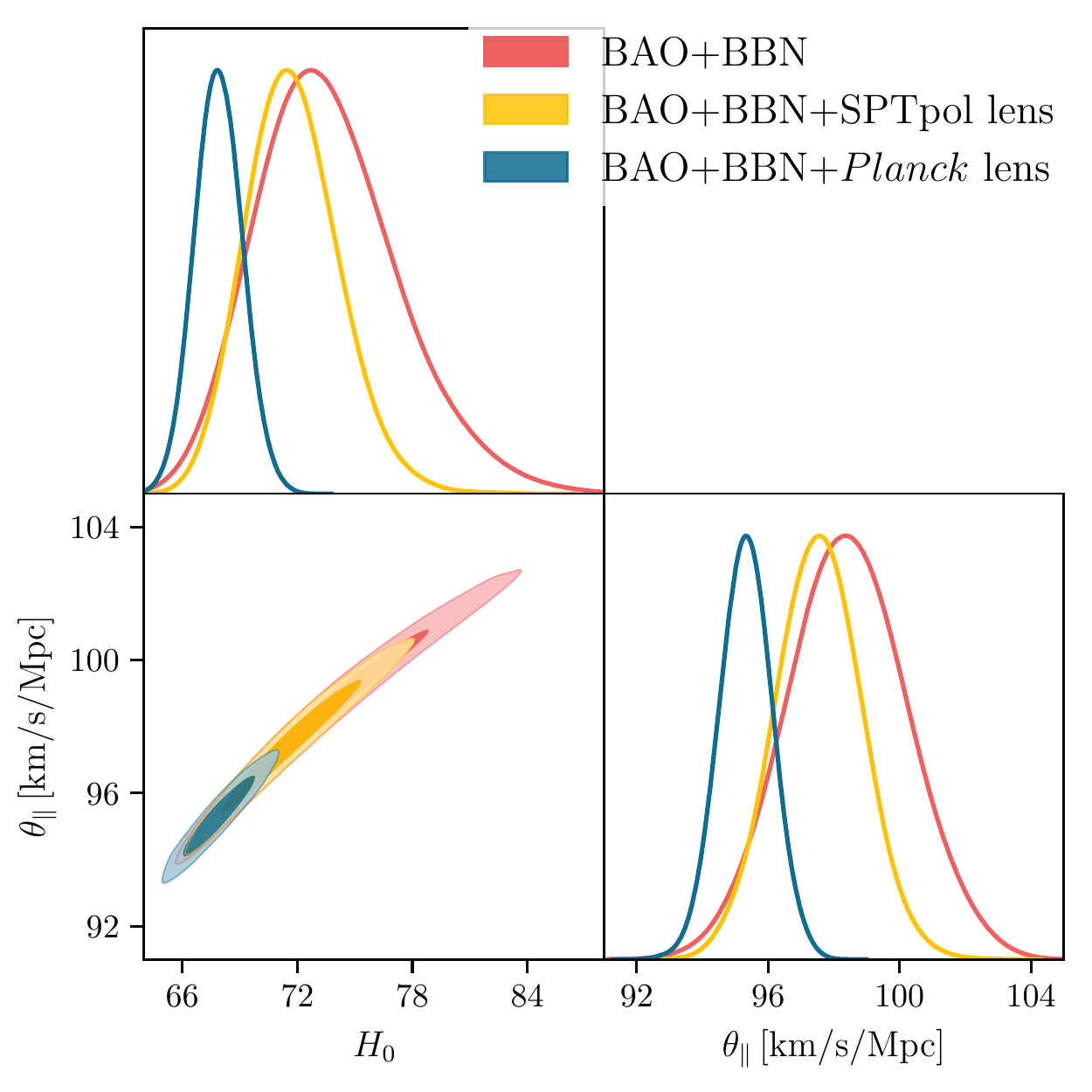}
\caption{The posterior distributions of the data sets 
\BB, \BBS, \BBP for $H_0$ and $\thetapar$.
The differences observed in $H_0$ are highly correlated with
the differences in $\thetapar$, which suggests a common origin in 
the parallel BAO measurements.
}
\label{fig:lens_bao_h0_rh}
\end{center}
\end{figure}

To tie the difference between \BBS and \BBP seen in $\thetapar$ to 
the difference originally identified in $H_0$,  
we show the posteriors of these two parameters in~\reffig{lens_bao_h0_rh}.
Note that the distributions and shifts in means follow each other due to the high correlation 
between the two parameters.   
We will hereafter use tension in $\thetapar$ as a proxy for this disagreement in $H_0$.
In the following sections, we use the $\QUDM$ analysis to quantify these tensions and further isolate their
origin in the various data sets.

\subsection{\BBS vs \planck\ lensing}

\begin{figure}
\begin{center}
\includegraphics[width=0.48\textwidth]{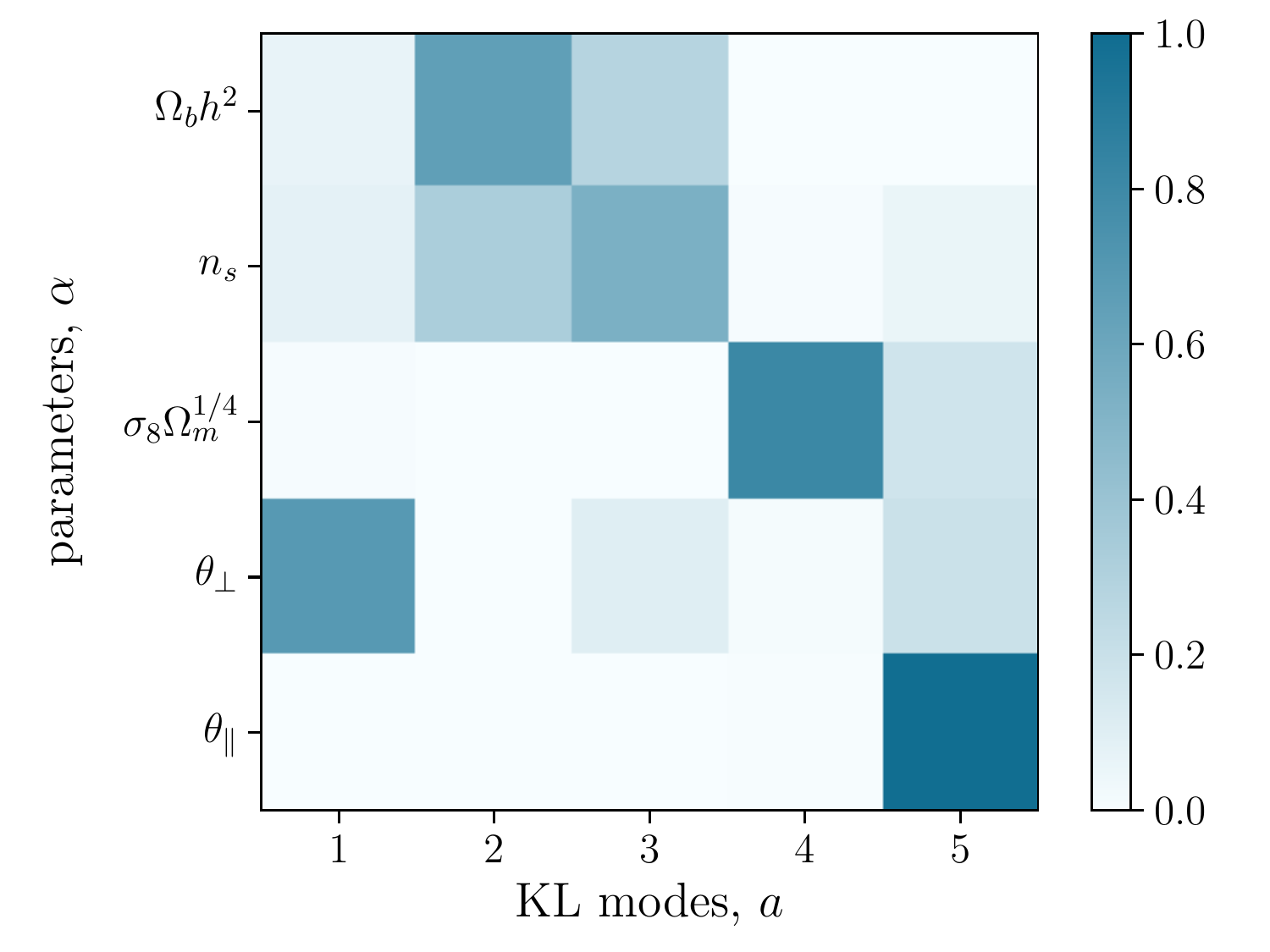}
\caption{Fractional Fisher information $F_{\alpha\alpha}^a / F_{\alpha\alpha}$
of the \mbox{\BBS} data set
computed using the KL eigenmodes from updating it
with
\planck\ lensing. 
The numbers in each row add to one.  The KL directions $a = 4$ and $a=5$ satisfy the KL update
criteria (Eq.~\ref{eq:KLrange}). 
They contribute the most information to the lensing amplitude parameter
$\sigma_8\Omega_m^{1/4}$ and
the parallel BAO parameter $\thetapar$ respectively.}
\label{fig:sptplk_fish_frac}
\end{center}
\end{figure}

We first analyze tension between \BBS and its update
BAO+BBN+\sptpol\ lensing+\allowbreak\planck\ lensing through the \QUDM\  statistic.
In this case, two directions satisfy the KL update criteria on eigenvalues  (Eq.~\ref{eq:KLrange}).
These two KL eigenmodes are $a=4,5$ and, as shown in \reffig{sptplk_fish_frac},
they dominate the Fisher information for the lensing amplitude parameter
$\sigma_8\Omega_m^{1/4}$ and the parallel BAO parameter $\thetapar$
respectively. 
We will refer to them below as the amplitude mode and the parallel mode.
With these two degrees of freedom, 
 $\QUDM = 4.0$, which corresponds to a 
probability-to-exceed (PTE) of $\sim$\,13\%.  We obtain similar values for all
DR12 values of $z_\BAO$.

Taken at face value, the PTE signals that the parameters of \BBS and
\planck\ lensing are not particularly in tension.
However, as can be seen in~\reffig{lens_bao_dm_h} and also~\cite{Bianchini:2019vxp}, 
the best-fit $\sigma_8\Omega_m^{1/4}$  between \sptpol\ lensing
and \planck\ lensing are almost too consistent with each other
despite being nearly uncorrelated in their lensing information.
It is therefore interesting to examine the individual contributions to $\QUDM$ of the
amplitude and parallel modes.  
We find that the amplitude mode contributes only $0.11$ whereas
the parallel mode contributes 3.9 to the total $\QUDM$.  
Considered separately, these correspond to a PTE of $74\%$ 
and $4.7\%$ respectively.  Because the contribution to $\QUDM$ from the amplitude  mode is smaller than expected,
the total significance downplays the tension in the parallel mode.
This parallel mode reflects the tension originally identified in $H_0$.  
Indeed if we computed the update difference in mean for only the marginal $H_0$ distributions,
we would obtain a 6.8\% PTE or effectively a 1.8\,$\sigma$ tension.

As discussed in the previous section, the \BB data do not contribute to the $\sigma_8\Omega_m^{1/4}$ constraint 
and the \sptpol\ lensing measurements contribute little to the $\thetapar$ constraint.
The $\QUDM$  contributions imply that the $\sigma_8\Omega_m^{1/4}$ measurements from \sptpol\ lensing and \planck\
lensing are slightly too consistent, while the $\thetapar$ parameter from the BAO data set
and \planck\ lensing are in mild disagreement of $\sim\,$2\,$\sigma$.
With this information, we now see that the difference 
between \sptpol\ lensing and \planck\ lensing when both are combined with \BB, first identified in $H_0$,
is actually driven by the mild disagreement in $\thetapar$ between \planck\ lensing and the BAO measurements.

We next focus on quantifying this disagreement between \planck\ lensing and \BB.

\subsection{\BB vs \planck\ lensing}
In the previous section, we determined that the parallel BAO parameter $\thetapar$ 
is the main indicator of tension in the data sets we consider.    Given that SPTpol lensing
has very little information on this parameter, we now focus on the comparison between
\BB and
\planck\ lensing.   

\begin{figure}
\begin{center}
\includegraphics[width=0.48\textwidth]{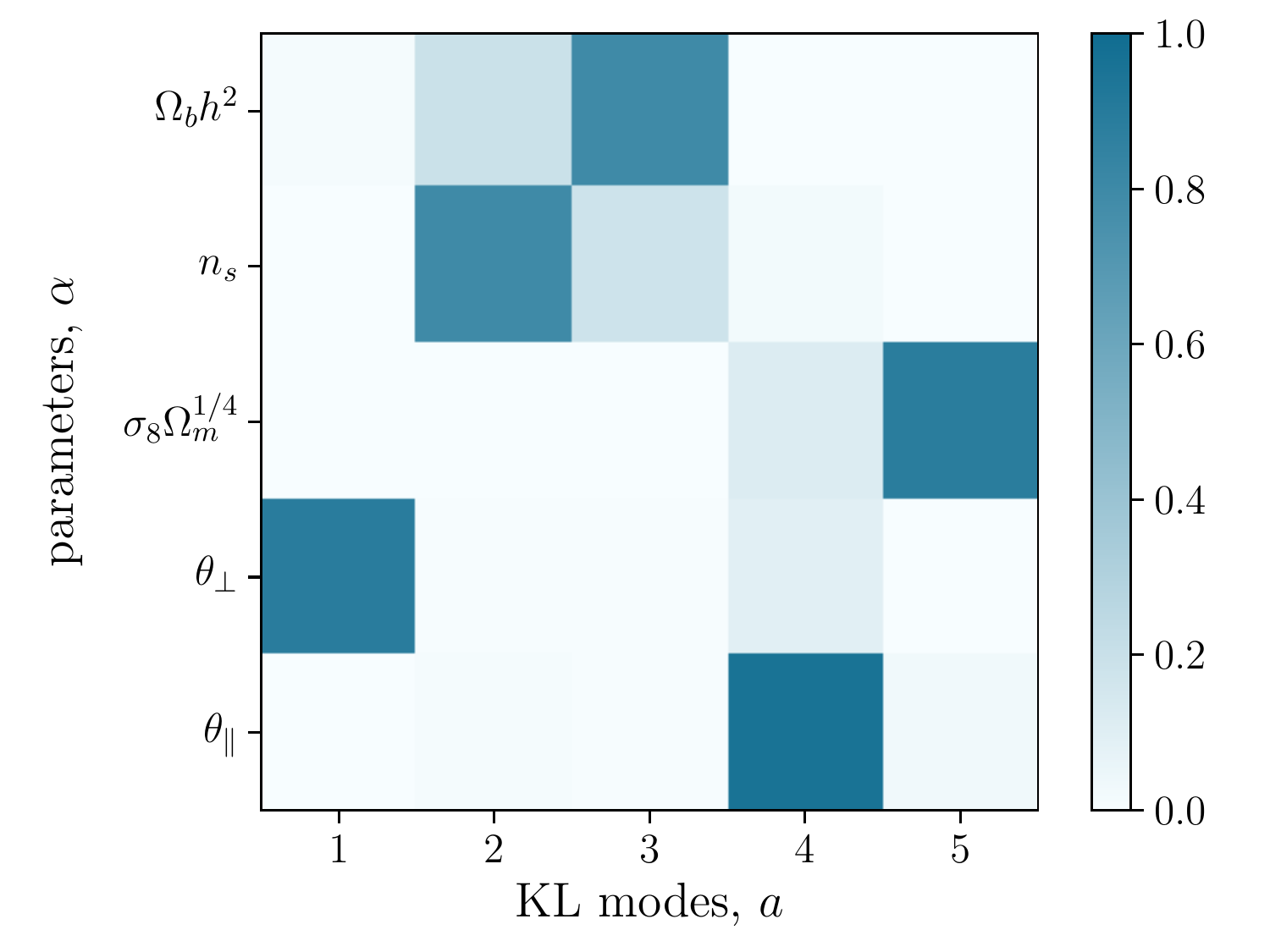}
\caption{Fractional Fisher information $F_{\alpha\alpha}^a / F_{\alpha\alpha}$
of the \BB data set
computed using the KL eigenmodes from updating it with  \planck\ lensing. 
Only the KL direction $a = 4$ satisfies the KL update criteria. It contributes the most information to
the parallel BAO parameter $\thetapar$.}
\label{fig:bao_fish_frac}
\end{center}
\end{figure}

In this case, we calculate  $\QUDM$ between \BB and its update \BBP. 
From the KL decomposition, only one mode satisfies the KL update criteria on
eigenvalues.
For \BB, this mode again dominates the information on  the parallel BAO parameter $\thetapar$, as shown
by the fractional Fisher information of  this ``parallel"  mode $a=4$ in~\reffig{bao_fish_frac}.  
In this case $a=5$ dominates the information in the lensing amplitude parameter but its constraints
come almost entirely from \planck\ lensing and so do not satisfy Eq.~(\ref{eq:KLrange}).

With the parallel mode,  $\QUDM=3.37$ for a single degree of freedom and 
hence the PTE is 6.6\%.    
We obtain similar values for all DR12 values of $z_\BAO$.
If we compute the update difference in means from the marginalized constraints on $H_0$ alone, we
obtain a PTE of 10.5\%, underestimating the significance.
To confirm that \sptpol\ lensing does not contribute to $\thetapar$ beyond what \planck\ lensing does
in this context, we calculate the $\QUDM$ between \BB and its update 
BAO+BBN+\planck\ lensing+\sptpol\ lensing. 
The eigenmodes have similar distributions as the \BB update with \planck\ lensing case and
the PTE is also 6.6\%, concluding that \sptpol\ lensing does not affect this result.
Again, but now more explicitly, the $\QUDM$ analysis shows that the parallel BAO parameter is in mild disagreement between
\planck\ lensing and \BB.

For the BAO data we have used in this work, the SDSS DR12 BAO data set have separate
parallel BAO and perpendicular BAO measurements. 
In the next section, we look into the effects on the BAO parameters from the parallel and the perpendicular
BAO measurements separately.

\subsection{BAO DR12 parallel vs \planck\ lensing}

\begin{figure}
\begin{center}
\includegraphics[width=0.48\textwidth]{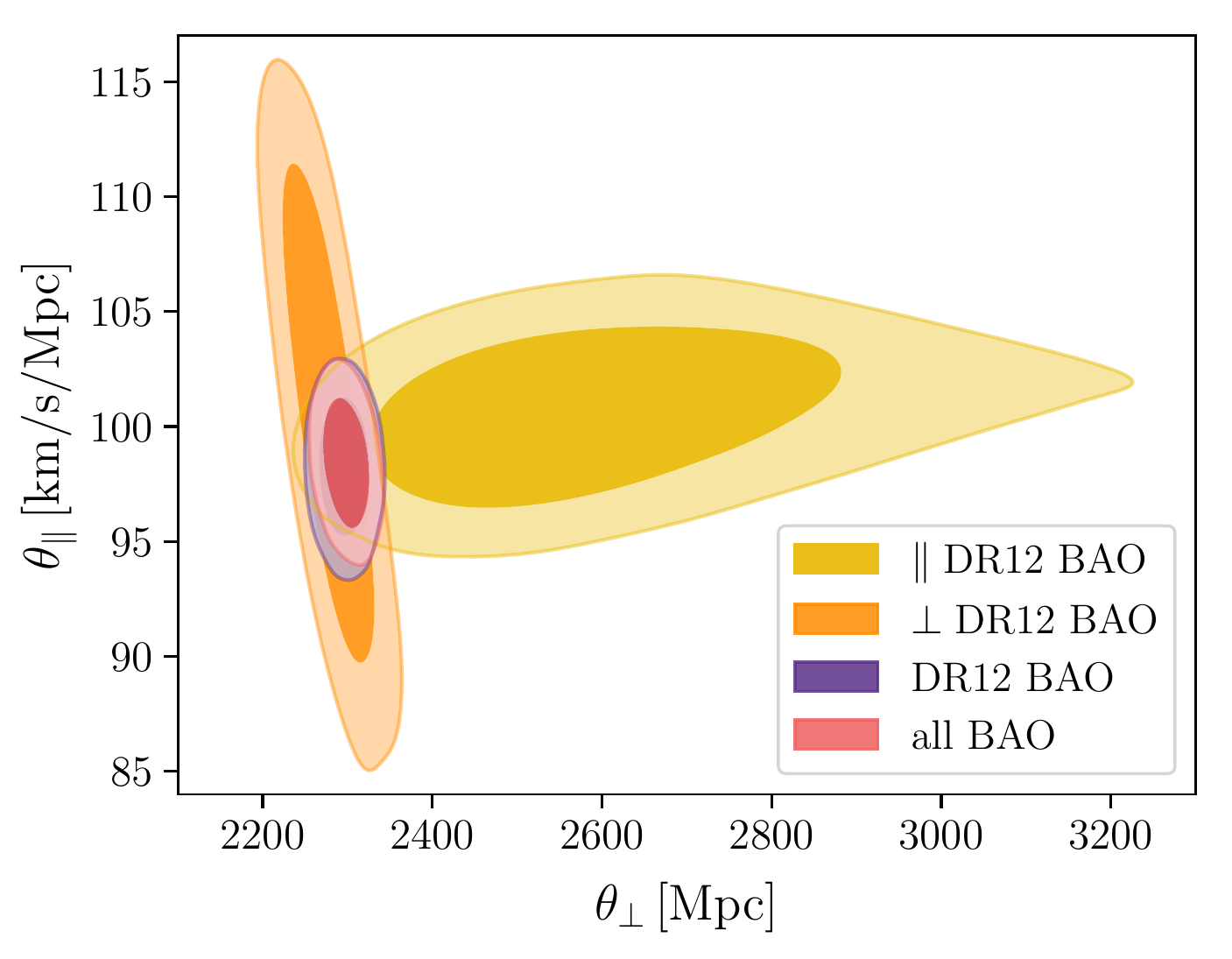}
\caption{$\thetapar$ vs $\thetaperp$ for various combinations of BAO data.}
\label{fig:bao_parallel_tang_2d}
\end{center}
\end{figure}

To identify the origin of the disagreement between \BB and \planck\ lensing,
we examine the posterior constraints on the  BAO parameters $\thetaperp$ and $\thetapar$ from
various subsets of the BAO measurements themselves  in~\reffig{bao_parallel_tang_2d}.  
Since these individual constraints are themselves too weak to have data-dominated 
Gaussian posteriors, we do not employ $\QUDM$ here.

First, we see that the BAO constraints mainly come from the DR12 points with little information from
other BAO data (MGS, 6dF).    
Next, we see that parallel and perpendicular DR12 measurements  map onto these  parameters without significant degeneracy, as expected given the 
design of the parameters.  
However, the correspondence between the measurements and parameters is not entirely one-to-one.
There is a small amount of constraining power of the perpendicular measurements on the
parallel  parameter and vice versa due to the combination of the three redshift points in each set.   
Recall that we take $\zBAO=0.61$ in our fiducial parameter choice, which is the highest of
the three DR12 redshifts.
We see in~\reffig{bao_parallel_tang_2d} that the 68\% CL regions of the two posteriors almost overlap,
which suggests that the parallel and perpendicular measurements might be in mild tension.
However, the level of tension cannot be inferred from this observation directly because of the correlation
between the parallel and perpendicular measurements.  
To properly account for the correlation, we apply the
parameter duplication technique \cite{Raveri:2019gdp} described in~\refsec{tensions}, and find that
the PTE associated with zero parameter difference is $32\%$, indicating no significant tension.

\begin{figure}
\begin{center}
\includegraphics[width=0.48\textwidth]{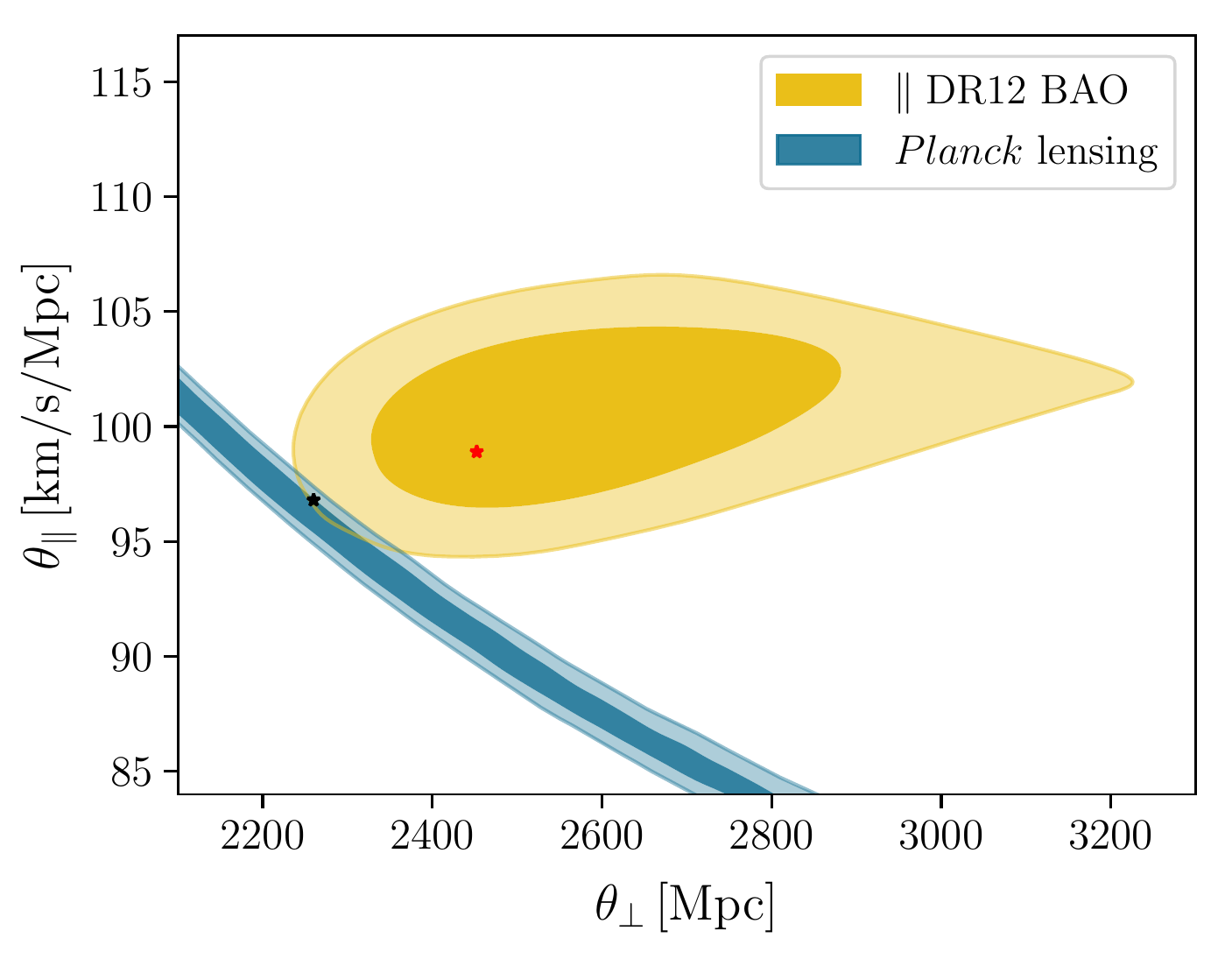}
\caption{$\thetapar$ vs $\thetaperp$  for parallel BAO and \planck\ lensing.
The red star is the best-fit model to the parallel BAO
measurements, the black star is a point we choose for
illustration. It lies on the
95\% CL line of the parallel BAO posterior and 68\% CL line of the \planck\ lensing
posterior.  We calculate the $\Delta\chi^2$ between these two models using
the parallel BAO likelihood. 
} 
\label{fig:bao_parallel_plklens_2d}
\end{center}
\end{figure}

Comparing the posteriors of the perpendicular and parallel BAO measurements with that
of \planck\ lensing in the $\thetapar-\thetaperp$ plane, we find good overlap between 
perpendicular BAO and \planck\ lensing but that the 95\% CL regions from parallel BAO
barely overlaps with the \planck\ lensing posterior, as shown in~\reffig{bao_parallel_plklens_2d}. 
However, we caution the reader that these two data sets have non-Gaussian posterior probabilities
and intuition of tension based on Gaussian data-dominated posteriors may not apply.
Specifically, due to the weak constraining power of the parallel BAO data in the $\thetaperp$ direction, 
the shape of the priors (assumed flat in $\theta_\mathrm{MC}, \Omega_c h^2$) over the
constrained range informs the shape of the posterior.

To demonstrate that the prior is informative, 
we consider a goodness-of-fit statistic for the parallel BAO data. We compare
$\chi^2_{\parallel\,\mathrm{BAO}}$ between the
best-fit parameters to the parallel BAO data (red star
in~\reffig{bao_parallel_plklens_2d}) and a representative test case
(black star). The latter sits on the 95\% CL line for the parallel BAO data,
which for a two dimensional Gaussian likelihood and flat priors in the BAO parameters would correspond to $\Delta \chi^2_{\parallel\,\mathrm{BAO}} = 6$. 
However, actually evaluating the
difference in $\chi^2_{\parallel\,\mathrm{BAO}}$ between the two cosmological models
highlighted in~\reffig{bao_parallel_plklens_2d} leads to only $\Delta
\chi^2_{\parallel\,\mathrm{BAO}} = 3.2$. Because of the increase of the prior volume at
high $\thetaperp$, 
which we shall see is associated with the large range in $\Omega_m$ that it encompasses,
the yellow contours in~\reffig{bao_parallel_plklens_2d} are shifted to the right compared to the position of the best-fit model.

This shows that, in fact, the two models are not as discrepant as one would normally infer
from a $95\%$ CL exclusion in two dimensions.  However, this mild discrepancy does account for a
large portion of the disagreement between  \BB and  \planck\ lensing in the previous section.  This is in
part because the \planck\ lensing constraint, and any tension with it, is effectively
one-dimensional in the BAO parameters, where a 95\% exclusion would correspond to $\Delta \chi^2 =3.8$.

\begin{figure}
\begin{center}
\includegraphics[width=0.48\textwidth]{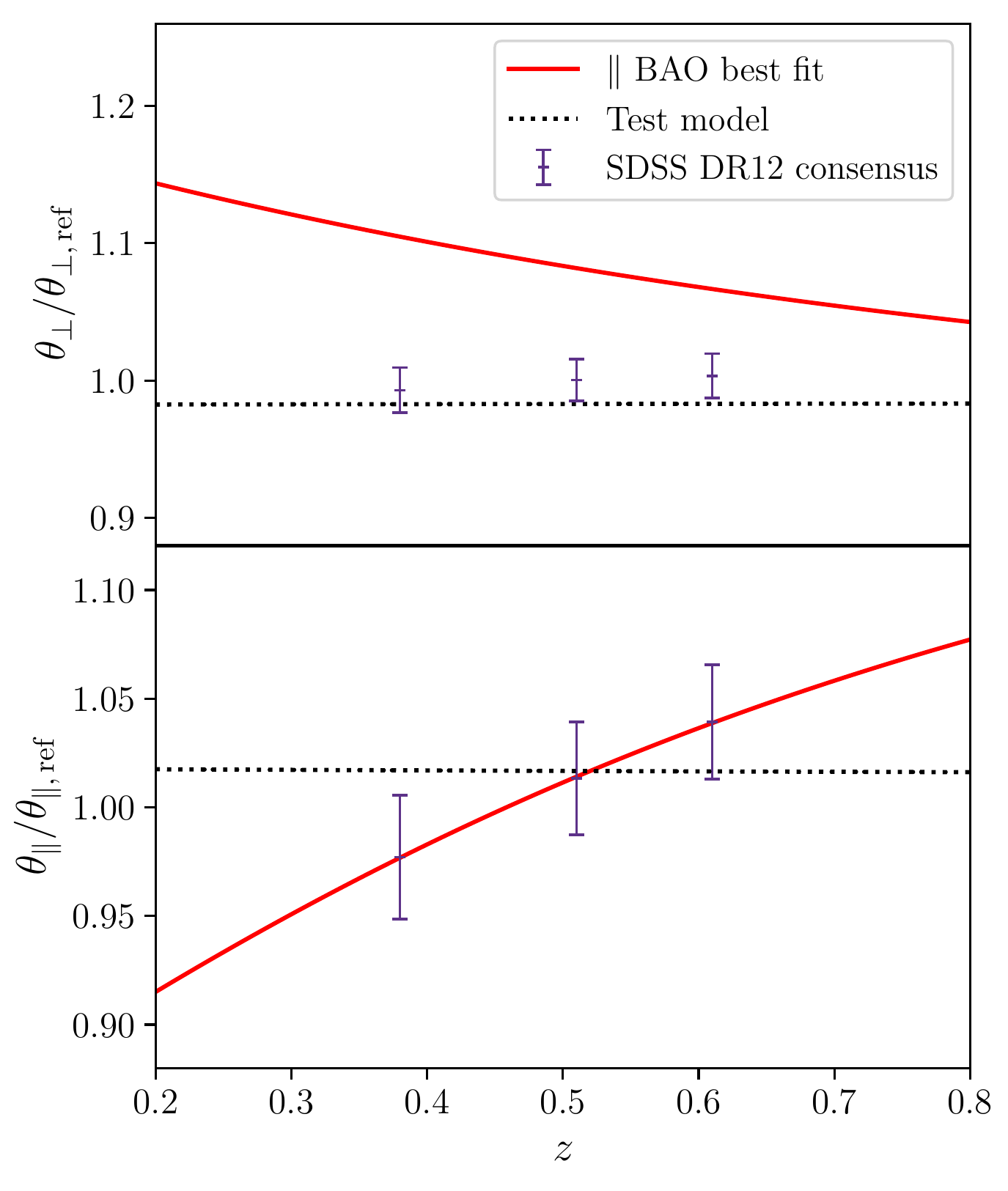}
\caption{SDSS DR12 BAO consensus measurements plotted against the $\thetaperp(z)$ and $\thetapar(z)$
model predictions from the parallel BAO best-fit 
and the test model, the red and black stars in~\reffig{bao_parallel_plklens_2d} respectively.
$\thetaperp$ and $\thetapar$ are normalized by $\theta_{\rm ref}$, the best-fit model to the \BBP data set. 
}
\label{fig:bao_data_bestfit}
\end{center}
\end{figure}

Correspondingly, we can trace these results back to the fits to the 3 redshift points of the
DR12 BAO data themselves.  We plot in~\reffig{bao_data_bestfit} the parallel and perpendicular
BAO measurements against the parallel BAO best-fit model and the test model, all relative to a reference model chosen as the best-fit to the  \BBP data set.
On the top panel, we see that the test model does not deviate from the \BBP best-fit very much. 
The parallel BAO best-fit model does not fit the perpendicular BAO measurements, with
$\chi^2$ of 57 for 3 data points.
On the lower panel, we show the parallel BAO measurements against the same sets of models.
The measurements are very well fit by the parallel BAO best-fit model.
The test model reflects the $\Delta
\chi^2_{\parallel\,\mathrm{BAO}} = 3.2$ shift noted above.
Moreover, we can now see that this penalty in the fit comes from its mismatch to the
 redshift slope of the parallel BAO data points.

\begin{figure}
\begin{center}
\includegraphics[width=0.48\textwidth]{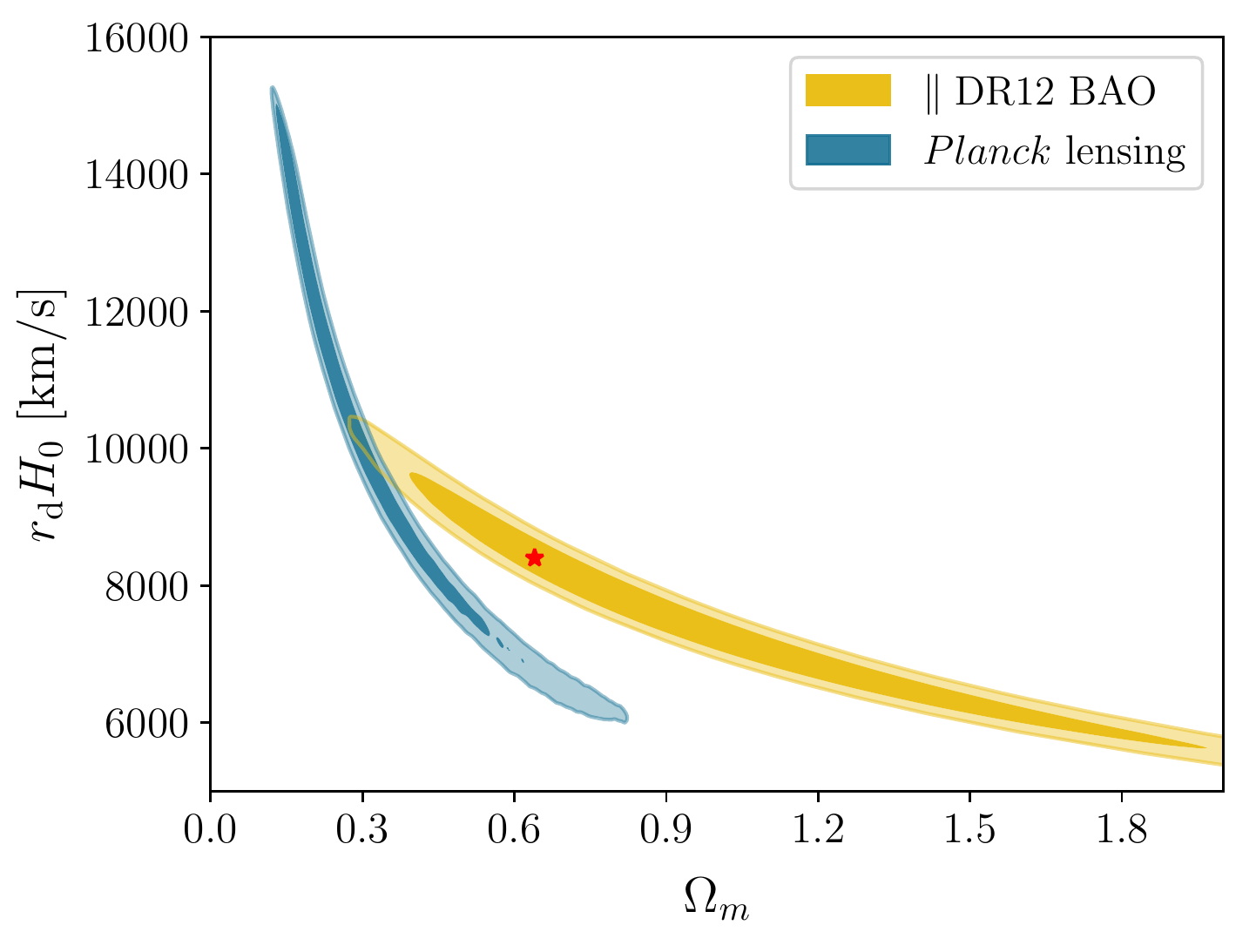}
\caption{
Constraints as in~\reffig{bao_parallel_plklens_2d} but here in terms of $\Omega_m$ and $\rdrag
H_0$. Note the long degeneracy for the parallel BAO data out to high $\Omega_m$ and the high value of
$\Omega_m=0.64$ for the parallel BAO best fit (red star).
}
\label{fig:om}
\end{center}
\end{figure}

Finally, to allow for easier comparison with literature and  with other
data sets in the next section, in Fig.~\ref{fig:om} we re-plot Fig.~\ref{fig:bao_parallel_plklens_2d} in the
parameter plane of $\Omega_m$ and $\rdrag H_0$. 
As already mentioned, these parameters have
 a one-to-one mapping with $\thetapar$ and $\thetaperp$.
From the figure, it is clear
that the parallel BAOs pull in the direction of very large $\Omega_m$.  
The best-fit parallel BAO model (red star) has $\Omega_m=0.64$ and $ \rdrag H_0 = 8400\, \mathrm{km/s}$. 
With the BBN prior, the best-fit also has a high $H_0= 74\, \mathrm{km/s/Mpc} $, but as we shall show in the next section, there are many other data sets that would exclude
the high $\Omega_m$ required.

\section{Comparison with other data sets}
\label{sec:otherdata}

\begin{figure}
\begin{center}
\includegraphics[width=0.48\textwidth]{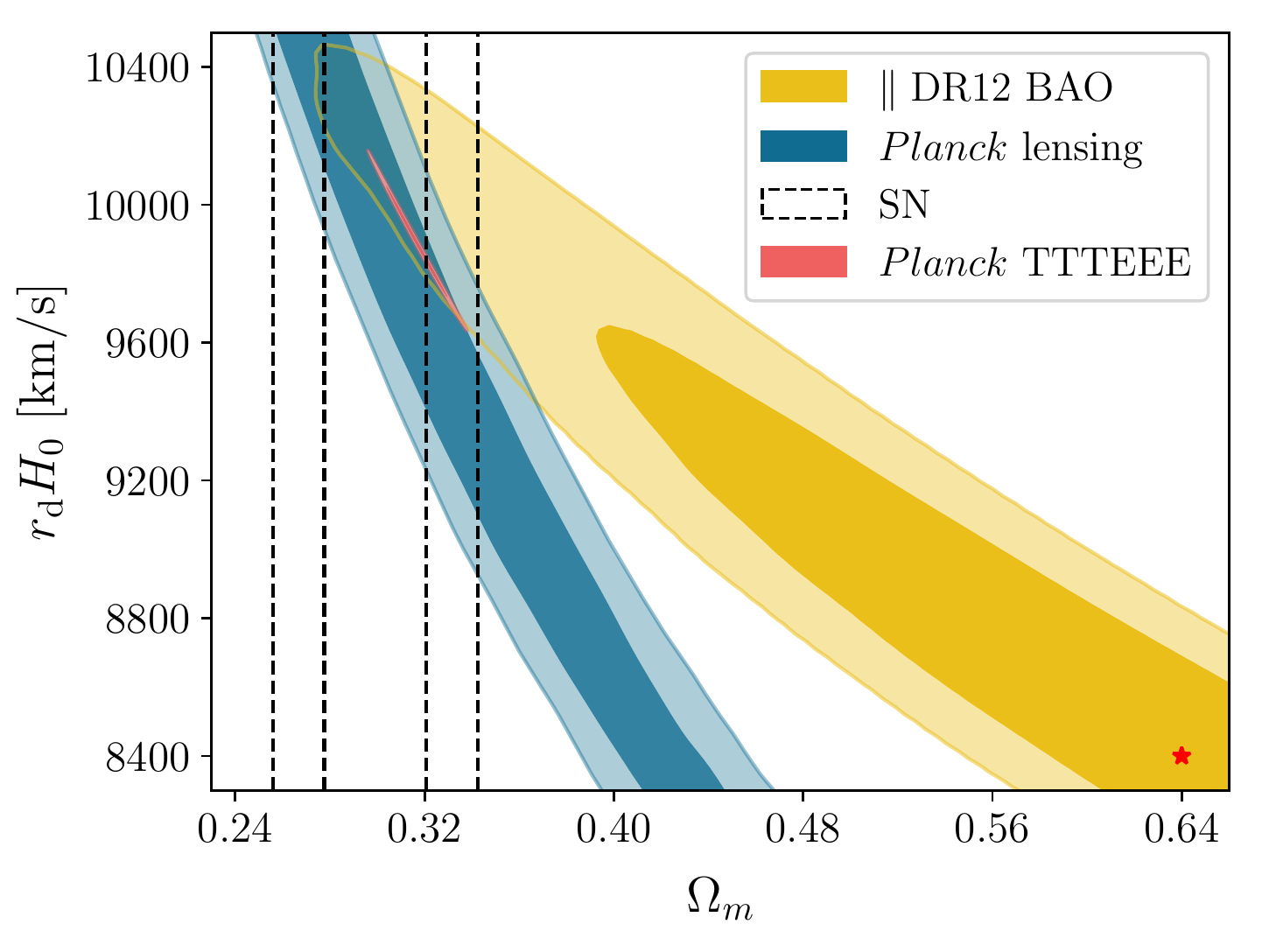}
\caption{
Constraints on $\Omega_m$ and $\rdrag H_0$ from parallel BAOs compared with those from
 \planck\ lensing, the Pantheon supernovae sample, and \planck\ CMB power
spectra.  Note that the posteriors extend beyond the part of
the parameter plane shown here.  
All of the other data sets strongly disfavor the parallel BAO best fit (red star)
and other models allowed by the parallel posterior at high $\Omega_m$.}
\label{fig:other_cosmo}
\end{center}
\end{figure}

To put things in context, in Fig.~\ref{fig:other_cosmo} we compare constraints on
$\Omega_m$ and $\rdrag H_0$ from the parallel BAO and \planck\ lensing measurements with those from
other cosmological data sets. 
We plot constraints on $\Omega_m$ from the Pantheon supernova sample
\cite{Scolnic:2017caz} assuming flat $\Lambda$CDM (dashed lines). While these supernova constraints are
in good agreement with \planck\ lensing and the combined BAO constraints, they are
in mild tension with the parallel BAOs. We then show constraints from 
the \planck\ primary CMB power spectra measurements \cite{Aghanim:2019ame} (red contours), which are compatible with  and even stronger than supernovae.

The best-fit $\Omega_m$ value of the parallel BAO+BBN data set (red star) is therefore strongly ruled out by 
both the supernova sample and the \planck\ primary CMB measurements.
We note that the higher redshift BAO measurements using Lyman-$\alpha$ forests or quasar clustering
also prefer low $\Omega_m$~\cite[e.g.][]{Addison:2017fdm, Blomqvist:2019rah, Agathe:2019vsu, Cuceu:2019for}.
This disallowed preference for high $\Omega_m$ in the parallel BAO data set is the ultimate origin of
 the high $H_0$ preferred by the \BBS data set compared with \BBP.

\section{Conclusion}
\label{sec:discuss}

In this work, we apply the update difference-in-mean statistic $\QUDM$ to quantify tension between 
two  composite data sets,  \BBS and \BBP, and track the origin of the differences in $H_0$ to the individual data sets that are
primarily responsible.
We work in a parameter basis that is native to the BAO and the CMB lensing measurements, replacing 
the cosmological parameters $\Omega_m$, $H_0$ and $A_s$ with  $\thetapar$, $\thetaperp$, and 
$\sigma_8\Omega_m^{1/4}$, where the parameter posteriors are nearly Gaussian and
 their constraints are relatively easy to map back to the measurements.
With this setup, we isolate the parameter direction that dominates the disagreement and isolate
its origin in the \planck\ lensing vs.~parallel BAO measurements.

We arrive at this conclusion through a process of narrowing down parameter combinations that matter and removing data sets
that contribute little to the disagreement.
In calculating the update difference in mean
 of \BBS updated by \planck\ lensing, the parameter direction that
dominates the tension is $\thetapar$ with a PTE of 4.7\%.   This direction is highly correlated with $H_0$, 
which carries a comparable tension.
Knowing that \sptpol\ lensing contributes little to $\thetapar$ constraints, we next check the $\QUDM$ of 
\BB updated by \planck\ lensing.
This test confirms the tension between \BB and \planck\ lensing along the $\thetapar$ direction
at a PTE of 6.6\%. 

Both update difference in mean statistics point to the parallel BAO parameter as the source of the disagreement.
We thus divide the BAO measurements into subsets to further our investigation. 
While the perpendicular BAO measurements are largely compatible with both the parallel BAO measurements and \planck\ lensing, there is disagreement
between the parallel BAO  and \planck\ lensing measurements around the 95\% CL.
This exclusion is exacerbated by our chosen prior, which in particular allows a large range  in $\Omega_m$.
Independent of this prior,  the $\Delta\chi^2$ between the best-fit model to the parallel BAO data and a representative
model that is consistent with \planck\ lensing
 is $\Delta\chi^2=3.2$ for effectively 1  degree of freedom.
These results indicate that the bulk of the difference between \planck\ lensing
and the BAO data set is indeed from the parallel BAO measurements.
Finally, we trace the origin of this $\Delta\chi^2$ to a slope in the parallel BAO measurements as 
a function of redshift, which drives its preference for high $\Omega_m$ values.  In combination with
constraints from BBN, this translates into a preference for high $H_0$ values.

We note that the $\Omega_m$ preferred by the parallel BAO data under $\Lambda$CDM is highly excluded
by other data sets, including supernova measurements and \planck\ primary CMB measurements. 
These other measurements tend to have much stronger constraining power on $\Omega_m$
than the parallel BAO data. 
For this reason, the mild disagreement of the parallel BAO data with the other data sets is hidden when 
analyzed in combination.   
In addition, the same holds when the galaxy BAO data are combined with higher-redshift Lyman-$\alpha$ 
BAO measurements~\cite{Aubourg:2014yra, Addison:2017fdm, Cuceu:2019for}.
While this tension clearly cannot be resolved within $\Lambda$CDM, it
is useful to bear this in mind when considering alternatives.

Finally, we reiterate the importance
of not selecting cosmological parameters {\it a posteriori} when adjudicating tension between data sets. 
Had we calculated the  difference in mean of \BBS updated with \planck\ lensing on $H_0$ alone, 
the PTE would be lower than letting the algorithm reveal that there are two relevant parameter combinations.
Conversely, had we calculated the  difference in mean of \BB updated with \planck\ lensing on $H_0$ alone, 
the PTE would be higher than letting the algorithm choose the single relevant parameter direction.
With parameters selected {\it a posteriori}, a trials factor is required to accompany the resultant PTE
for fair interpretation of the statistic and that selection may still not reflect the true source of tension.
Looking forward, as upcoming surveys provide more precise measurements of our universe, it is of utmost
importance that we identify the origin and significance of tension accurately to aid the differentiation of the underlying
causes of the observed tension---be it unmodeled systematics or new physics.

\acknowledgements  We thank Georgios Zacharegkas for useful discussions.  
WLKW is supported in part by the Kavli Institute for Cosmological Physics at the University of Chicago through grant NSF PHY-1125897 and an endowment from the Kavli Foundation and its founder Fred Kavli.
WH is supported by by U.S.~Dept.~of Energy contract DE-FG02-13ER41958 and the Simons Foundation. 
MR is supported in part by NASA ATP Grant No. NNH17ZDA001N, and by funds provided by the Center for Particle Cosmology. 
This work was completed in part with resources provided by
the University of Chicago Research Computing Center.

\bibliography{biblio}
\end{document}